\input{aipcheck}

\documentclass[
    ,final            
  ]
  {aipproc}

\layoutstyle{6x9}

\begin{document}

\title{Gauge invariance and lattice monopoles}


\author{A. Di Giacomo}{
  address={Dipartimento di Fisica and INFN, Pisa, Italy}
}

\begin{abstract}
 The number and the location of monopoles
 in Lattice configurations depend on the choice of the gauge,  in contrast to the obvious requirement that monopoles, as 
physical objects,  have a gauge-invariant status.
  It is proved, starting from non-abelian Bianchi identities, that monopoles are
indeed gauge-invariant: the technique used to detect them has instead an 
efficiency which depends on the choice of the abelian projection, in a known and 
well understood way.

 \end{abstract}

\maketitle


\section{Introduction}

 Monopoles are relevant excitations of $QCD$: in particular they can condense in the vacuum and produce dual superconductivity, which is a candidate mechanism for confinement \cite{'tHP}\cite{m}\cite{'tH2}.
 If dual superconductivity is the mechanism of confinement monopoles are expected to dominate the
 critical dynamics at the deconfinement phase transition. Evidence is claimed that they also dominate
 below it, down to zero temperature\cite{suz}\cite{pol}.
 
 On the lattice monopoles have been studied from two complementary perspectives
 \begin{itemize}
 \item
 The symmetry of the vacuum state: an order parameter is introduced, which is the vacuum expectation value ($vev$) of a magnetically charged operator, and can discriminate between dual superconducting and normal phase\cite{digz,dp,ddp,ddpp,dlmp,ccos,3,vpc}.
 \item
 The observation of monopoles in lattice configurations, trying to extract a monopole effective action, in particular to check monopole dominance\cite{sco},\cite{suz,pol}.
 \end{itemize}
 
 In this paper we shall concentrate on the second approach.
 
 The procedure to detect monopoles in lattice configurations was first developed in Ref.\cite{dgt} for $U(1)$ gauge theory. Any excess over $2\pi$ of the abelian phase of a plaquette is interpreted as existence of a Dirac string through it, and a monopole exists in an elementary cube whenever a net number of Dirac flux-lines crosses the plaquettes at its border. The phase of a $U(1)$ plaquette is gauge invariant, and hence the procedure is well defined and unambiguous. For a non abelian gauge theory
 one has first to fix a gauge \cite{'tH2} and then apply the same procedure to the components along an abelian subgroup of the gauge group (Abelian projection). The result is strongly gauge dependent and
 because of that the existence of a monopole in a given location of a configuration is a gauge dependent property. That is of course physically unacceptable.
 
 The prototype magnetically charged configuration is the soliton solution of Ref.s\cite{'tH,Pol} of an
 $SU(2)$ gauge theory coupled to a Higgs field in the adjoint representation.  In that solution the magnetic $U(1)$ which couples to the magnetic charge coincides with the little group of the $vev$ in the unitary representation of the Higgs field which produces the symmetry breaking. The magnetic charge corresponds to a non-trivial homotopy, the solution being a mapping of the sphere $S_2$ at spatial infinity on $SU(2)/U(1)$. This feature is gauge invariant.
 
 In $QCD$ there is no Higgs field, but the magnetic $U(1)$ has to be a suitable subgroup of the gauge group. It follows from general arguments that the magnetic-monopole term in the multipole expansion of any field configuration is intrinsically abelian and obeys abelian equations of motion \cite{coleman}. Since at a superficial sight there is no preferred gauge direction in $QCD$ it was proposed in Ref.\cite{'tH2} that any operator in the adjoint representation could be used as an effective Higgs field to identify the magnetic $U(1)$ subgroup, physics being for some reason independent on that choice (Abelian projection). It was soon realized that the location and the number of monopoles in a given configuration was projection dependent, and most people did choose the so called maximal-abelian gauge as the most suitable  for monopole dominance. In fact the number and the location of monopoles strongly depends on the choice of the abelian projection. 
 
  Can  monopoles be defined in a gauge-invariant way?
 
 We will show that this is possible, and that the difficulties outlined above are only apparent.
 By use of the Non Abelian Bianchi Identities ($NABI$), we will show that for any magnetically charged gauge field  configuration there exists a privileged direction in color space which identifies the magnetic $U(1)$.
 
 As a consequence monopoles have a gauge invariant physical status, as they should. If, however, one tries to detect them by fixing an abelian projection and by looking for Dirac strings with the technique of Ref.\cite{dgt} the result is projection dependent. In the maximal abelian projection that technique works 
 and monopoles are detected. In other abelian projections a fraction of the monopoles miss detection,
 and in the Landau Gauge projection no monopole is observed.  Dual superconductivity, instead, defined as Higgs breaking of the residual $U(1)$ symmetry is a gauge invariant feature of the system.

\section{The non abelian Bianchi identities.}

The abelian Bianchi Identities are the homogeneous Maxwell's equations
\begin{equation}
\partial_{\mu} F^*_{\mu \nu} =0 
\end{equation}
or  $\vec \nabla \cdot \vec B =0$ and $\vec \nabla \wedge \vec E + \partial _{t} \vec B=0$.

A violation of the Bianchi identities 
\begin{equation}
\partial_{\mu} F^*_{\mu \nu} = j_{\nu} \label{abi}
\end{equation}
means existence of a non-zero magnetic current $j_{\nu}$, which is conserved because of the antisymmetry of the tensor $F^*_{\mu \nu}$, 
\begin{equation}
 \partial_{\nu} j_{\nu} =0
 \end{equation}

The non abelian counterpart of Eq.(\ref{abi}) is
 \begin{equation}
 D_{\mu} G^*_{\mu \nu} = J_{\nu} \label{nabi}
 \end{equation}
 where $D_{\mu}$ denotes  covariant derivative, $G^*_{\mu \nu}$ is the dual of the non abelian field strength
 $G_{\mu \nu}$,  $G^*_{\mu \nu}= \frac{1}{2} \epsilon_{\mu \nu \rho \sigma} G_{\rho \sigma}$.
 
 The non-abelian current is covariantly conserved
 \begin{equation}
 D_{\mu} J_{\mu} =0
 \end{equation}
 as a consequence of Eq.(\ref{nabi}).
 
 The four components of the current  $J_{\mu}$ commute with each other, as always when the algebra of the generators of the symmetry does not involve generators of the Poincare' group \cite{cm}

Eq.(\ref{nabi}) is gauge covariant. To extract the gauge invariant information contained in it one can go to the representation in which the currents are diagonal and project on a complete set of diagonal matrices.
There are $r$ of them, if $r$ is the rank of the gauge group, by definition of rank. One possible choice are the fundamental weights $\phi^a_{0}$ ($a = 1,..,r$). There exists one fundamental weight for each
simple root of the group algebra $\vec \alpha^a$, which commutes with the  Cartan elements 
of the algebra $H_{i}$, $(i=1,..r)$ , $[\phi^a_{0}, H_{i}] =0$ and has commutators with the operators $E_{\pm \vec \alpha}$ related to the roots $\vec \alpha$ $[\phi^a_{0}, E_{\pm \vec \alpha}] = \pm (\vec c^a\cdot \vec \alpha)  E_{\pm \vec \alpha} $ . In particular for the simple roots  $\vec c^a\cdot \vec \alpha^b  = \delta_{ab}$,
so that, if we define $T^a_{3} \equiv \frac{1}{2} [ E_{ \vec \alpha^a}, E_{- \vec \alpha^a}]$, 
$Tr (\phi^a_{0} T^b_{3}) = \delta _{ab}$. 

If we denote by $\phi^a_{I} $ the matrix transforming in the adjoint representation which coincides with
$\phi^a_{0}$ in the representation in which $J_{\mu}$ is diagonal , the projection of Eq.(\ref{nabi}) described above 
reads
\begin{equation}
Tr\left ( \phi^a_{I} D_{\mu} G^*_{\mu \nu}\right) = Tr\left ( \phi^a_{I}J_{\nu}\right) \equiv j_{\nu}(I)
\end{equation}
For reasons which will be clear in the following we can also operate the projection on the matrix
$ \phi^a_{V}$ which, in the representation in which $J_{\mu}$ is diagonal, is defined as

$ \phi^a_{V} =  V(x) \phi^a_{0} V(x)^{\dagger} $

with $V(x)$ an arbitrary gauge transformation, getting
\begin{equation}
Tr\left ( \phi^a_{V} D_{\mu} G^*_{\mu \nu}\right) = Tr\left ( \phi^a_{V}J_{\nu}\right) \equiv j_{\nu}(V) \label{pr}
\end{equation}

 We shall denote by $F^a_{\mu\nu}(V)$ the 'tHooft tensor in the abelian projection in which $\phi^a_{V} $
 is diagonal, i.e. the tensor which coincides with the abelian field strength of the residual $U(1)$ symmetry in the gauge in which $\phi^a_{V} $ is diagonal.
 
 In Ref.\cite{bdlp} we have proved the following theorem which is valid for a generic compact gauge group.
 
{\bf THEOREM. As a consequence of Eq.\ref{nabi} for any compact gauge group Eq.\ref{pr} is equivalent to 
\begin{equation}
\partial_{\mu} F^{a*}_{\mu \nu}(V) = j_{\nu}(V) \label{th}
\end{equation} }

The breaking of the abelian Bianchi identities, i.e. the magnetic currents of the residual $U(1)$ gauge field in any abelian projection, are the projections on the corresponding fundamental weight of the
currents which break the non abelian Bianchi identities.

Eq.(\ref{th}) will be our tool to solve the problem we are concerned with. 

It implies
$\partial_{\nu} j_{\nu}(V) =0$
for arbitrary $V$. 

\section{The 't Hooft Polyakov  monopole revisited}

We shall now check our theorem on the soliton configuration of Ref.s \cite{'tH,Pol}. The configuration is a static solution of the SU(2) Higgs model, with the Higgs field in the adjoint representation. The Lagrangean has the form
\begin{equation}
L = -\frac{1}{4}\vec G_{\mu \nu}\vec G_{\mu \nu} + (D_{\mu} \phi)^{\dagger}(D_{\mu} \phi) -V(\phi^2)
\end{equation}
The notation is standard. The potential $V(\phi^2)$ contains a term linear in $\phi^2$ which has a negative $(-m^2)$ coefficient in the Higgs broken phase where the soliton exists, and a quadratic term in 
$(\phi^2)^2$ with positive coefficient $\lambda$.

The solution is worked out in the so called hedgehog gauge, in which the Higgs field at the position $\vec r$ in physical space is directed as $\hat r$ in color space. In formulae
\begin{equation}
\vec \phi(\vec r) = H(r) \hat r \label{hedg}
\end{equation}
with $H(r)_{r \to \infty} \to v$ the vacuum expectation value of $\phi$
 Eq.(\ref{hedg}) explicitly shows the non trivial mapping of the $S_{2}$ sphere at spatial infinity on $SU(2)/U(1)$.

The solution reads \cite{'tH,Pol}
\begin{eqnarray}
\vec A_{0}& =&0   \\
A^a_{i}& = &\epsilon_{iak}\frac{r^k}{gr^2}[1 - K(gvr)]
\end{eqnarray}
The function $K(x)$ depends on the parameters of the potential $V(\phi^2)$ but generically, modulo possible logs decays exponentially at large $r$ $K(x)_{x\to \infty} \propto \exp(-x)$ and, at small distances behaves as $[1 - K(x)]_{x\to0} \propto x^2$.
It is trivial to realize, by inspecting the solution, that this gauge is nothing but the Landau gauge
\begin{equation}
\partial_{\mu} A_{\mu}=0
\end{equation}
The 'tHooft tensor, or better the field strength along the diagonal component $\sigma_3$ can explicitly be computed getting for the abelian magnetic field
\begin{equation}
\vec b \approx _{r\to \infty} \frac{2 \hat r}{gr^2} \frac{z}{r}
\end{equation}
The magnetic charge $Q_{m}$ computed as flux of the magnetic field through the sphere at infinity
is trivially zero in this gauge! $Q_{m} =0 $ in Landau gauge.

Let us now go to the unitary gauge in which the Higgs field is rotated to a fixed direction in color space, say the $3-$axis. The configuration can easily be computed (see e.g. the appendix of Ref.\cite{shnir})
and it is trivial to verify that this gauge is nothing but the maximal abelian gauge defined by the condition\cite{'tH2}
\begin{equation}
\partial_{\mu}A_{\mu}^{\pm} \pm ig \left[A_{\mu}^3, A_{\mu}^{\pm} \right] =0
\end{equation}
Moreover in this gauge the non abelian magnetic current is diagonal, so that it coincides with the one identified by $\phi^a(I)$ of Sect.2.
In this gauge the magnetic charge can be computed either directly, as in the Landau gauge, or by our theorem. 

The solution being static the space components $J_{i}$ of the non abelian magnetic current vanish,
and the temporal component is directly computed to be
\begin{equation}
J_{0}  = D_{i}B_{i} = \frac{2\pi}{g} \delta^3(\vec r) \sigma_{3}
\end{equation}
There is one fundamental weight ($r=1$ for $SU(2)$) namely $\phi_{0} = \sigma_3$ . Our theorem Eq.(\ref{th})
gives
\begin{equation}
\vec \nabla\cdot \vec b= Tr(\phi_{0} J_{0}) = \frac{4\pi}{g}\delta^3(\vec r)
\end{equation}
or  $Q_{m} =\frac{1}{g} $ in the maximal abelian gauge.

If our soliton were a lattice configuration and we had measured the magnetic charge as the flux at infinity of the abelian field in the $3-$direction, as is usually done on the lattice, we would have found  a monopole of charge 2 Dirac units in the maximal abelian gauge, and no monopole at all in the Landau gauge.
In both cases the monopole would be there, and with it the magnetic charge which is determined  by the homotopy of the solution. It is not true that all abelian projections are equivalent. 
 The monopole has a preferred direction in color space, which is, in this case, the direction of the Higgs breaking, which coincides with the direction of the magnetic field at large distances in the unitary representation.
 \section{General case}
 The argument can be extended to a generic static configuration by use of a theorem due to Coleman[\cite{coleman} ,Sect. 3.3] 
 
{\bf The magnetic monopole term in the multipole expansion of a generic static field configuration is abelian : it obeys abelian equations of motion and can be gauged along one direction in color space, modulo a global transformation.}

If the configuration is not static one can apply the argument to the superposition of it to the time-inverted configuration to isolate the magnetic field due to monopoles at large distances.

Going to the gauge in which the asymptotic magnetic field is directed along a given direction in color space, say the $3-$axis, the non abelian magnetic field at large $r$ is fixed by the total magnetic charge $m$, to be
\begin{equation}
\vec B = \frac{m}{2} \frac{\vec r}{2gr^3}\sigma_3
\end{equation}
Since non diagonal terms in $\sigma_{\pm}$ are non leading it can easily be proved that also the
leading term of the abelian magnetic field at large distances is fixed.
\begin{equation}
\vec b \approx_{r\to\infty} \frac{m}{2} \frac{\vec r}{2gr^3}
\end{equation}
The gauge field at large distances obeys the maximal abelian gauge condition. 

In the maximal abelian gauge, or in any gauge which differs from it by an arbitrary gauge transformation $W(\vec r)$ which tends to the identity as $r \to \infty$, the flux of the abelian magnetic field at large distances directly gives the correct magnetic charge.

\section{Monopole condensation and confinement.}

The magnetic charge density in the maximal abelian projection is given by
\begin{equation}
j_0(x, I) = Tr(\phi_I J_{0}(x) )
\end{equation}
with $J_{0}$ the violation of the NABI Eq.(\ref{nabi}), and $\phi_{I}$ the fundamental weight diagonal with it.  The equal-time commutator of $j_{0}(x, I)$ with any local operator $O(y)$ carrying magnetic charge 
$m$ is
\begin{equation}
\left[ j_0(\vec x, x_0,I), O(\vec y, x_0)\right] = m \delta^3(\vec x -\vec y)O(\vec y,x_0)  + S.T.
\end{equation}
 By $S.T.$ we mean Schwinger terms.
 After integration over $\vec x$ the Schwinger terms give zero contribution and
 \begin{equation}
\left[Q(I),O(y)\right] = m O(y)
\end{equation}
If $m\neq 0$ and $\langle O \rangle \neq 0$ the magnetic $U(1)$ is Higgs-broken, and there is dual superconductivity.
In a generic abelian projection the magnetic current is
\begin{equation}
j_{0}(x, V) = Tr( V(x)\phi_{I} V^{\dagger}(x) J_{0})
\end{equation}
 $j_{0}(x, V)$ is gauge invariant, but we can compute it in the gauge in which $J_{0}$ is diagonal and $\phi_{I}$ with it. Since $V(x) \phi_{I} V^{\dagger}(x)$ belongs to the algebra it will be
 \begin{equation}
 V(x) \phi_{I} V^{\dagger}(x) = C(x,V) \phi_{I} +\sum  _{\vec \alpha} E_{\vec \alpha} D^{\vec \alpha}(x,V) \label{CCC}
 \end{equation}
 We have written the expansion having in mind the gauge group $SU(2)$ for simplicity. In the generic case the diagonal term will be, instead of a coefficient times $\phi_{I}$ a combination of the fundamental weights $\phi_{I}^a$ $( a= 1,..r)$, and the argument would be analogous.  In the gauge chosen only the first term contributes and 
 \begin{equation}
 \left[Q(V),O(y)\right] = m O(y)C(y,V)
 \end{equation}
  Since $C(y,V)$ is generically non-vanishing  the operator $O$ will have a non zero charge also
  in the new abelian projection, and if   $\langle O \rangle \neq 0$ also the new $U(1)$ is Higgs broken.
  Dual superconductivity is a gauge invariant physical property. This is in agreement with the numerical results obtained in Ref.\cite{suzu}.
  
  \section{Lattice Monopoles.}
  The recipe to detect monopoles \cite{dgt} on the lattice is based on the measurement of the abelian magnetic flux in a given abelian projection through the boundary surface of elementary cubes.
  The above analysis shows that the magnetic flux is equal to the "true" magnetic flux in the maximal abelian projection, or in any other projection which differs from it by a continuous gauge transfomation which tends to the identity at large spatial distances. In the procedure it is assumed that the border of the elementary cube is far enough to be at infinity. The choice of the gauge on the lattice is then limited to the maximal abelian gauge itself, modulo a check that, defining the flux by larger cubes leaves
  the flux unchanged. This proves to be the case in the maximal abelian gauge \cite{ddmo}\cite{bdd}
In other abelian projections the magnetic flux is expected to be smaller then the true one\cite{bdlp}.
This can be directly checked as follows\cite{bdd}:
\begin{itemize}
\item Fix the gauge to be the maximal abelian : monopoles appear as elementary cubes with a net number of Dirac strings crossing the border. In the weak coupling regime, near the continuum limit, 
almost all of them will have one single string attached. 
\item Assume as a first approximation that the monopole is in the centre of the cube and that the Dirac string is perpendicular to the plaquette crossed.
\item Perform gauge transformations depending on one parameter ${\bf a}$ $(0 \le {\bf a} \le1)$
of the form
\begin{equation}
 U({\bf a}) = \exp(- i\phi\frac{\sigma_3}{2}) \exp(- i{\bf a}\theta \frac{\sigma_2}{2}) \exp( i\phi\frac{\sigma_3}{2})
 \end{equation}
 Here $\theta$ and $\phi$ are the polar angles with respect to the direction of the Dirac string.
 For ${\bf a}=0$ $U({\bf 0})=1$ and one stays in the maximal abelian gauge. For  ${\bf a}=1$ $U({\bf 1})$
 is nothing but the unitary transformation bringing from the maximal abelian to the Landau gauge \cite{shnir}.  One can compute analytically the magnetic charge (flux of the abelian magnetic flux at large $r$) as a function of ${\bf a}$ , getting
 \begin{equation}
 \frac{Q({\bf a})}{Q({\bf 0})} = \frac{1 + \cos({\bf a}\pi)}{2}
 \end{equation}
 to be compared with lattice measurements. The comparison was done in Ref.\cite{bdd} after the time when this talk was presented, and is surprisingly positive, in spite of the approximations quoted above and of the discretization errors.
 \end{itemize}
 This confirms that the approach is correct and that the differences between abelian projections are well understood.
 \section{Concluding remarks}
 \begin{itemize}
 \item Monopoles are defined by their topological structure (homotopy), which is gauge invariant. They are intrinsically abelian entities. 
 \item Each field configuration with non zero magnetic charge has an intrinsically built in preferred 
 direction in color space, related to the magnetic monopole term in the multipole expansion of the 
 field at large distances, which is the natural $U(1)$ coupled to  monopole charge.
  This direction is selected by the so called maximal abelian gauge. In this gauge the recipe of Ref.\cite{dgt} really detects monopoles, defined as configurations with non trivial homotopy. 
    \item   The magnetic charge of the same configuration measured by different abelian projection,
  is projection(gauge) dependent. This explains why the location and the number of monopoles can be  different in different gauges. The recipe of Ref.\cite{dgt} in gauges other than the maximal abelian can miss monopoles.
  \item  Non abelian Bianchi identities allow to relate magnetic charges in different abelian projections.
  \end{itemize}
  \section{Aknowledgements}
  The author is grateful to Claudio Bonati, Massimo D'Elia, Luca Lepori and Fabrizio Pucci,who collaborated to  the works Ref.\cite{bdlp}, \cite{bdd} where the physics presented this talk
  was developed.




\bibliographystyle{aipproc}   

\bibliography{sample}



\end{document}